# Optimized Disaster Recovery for Distributed Storage Systems: Lightweight Metadata Architectures to Overcome Cryptographic Hashing Bottleneck

Prasanna Kumar, Nishank Soni, Gaurang Munje

*Abstract*—Distributed storage architectures are foundational to modern cloud-native infrastructure, yet a critical operational bottleneck persists within disaster recovery (DR) workflows: the dependence on content-based cryptographic hashing for data identification and synchronization. While hash-based deduplication is effective for storage efficiency in steady-state operation, it becomes a systemic liability during failover and failback events when hash indexes are stale, incomplete, or must be rebuilt following a crash. This paper precisely characterizes the operational conditions under which full or partial re-hashing becomes unavoidable. The paper also analyzes the downstream impact of cryptographic re-hashing on Recovery Time Objective (RTO) compliance, and proposes a generalized architectural shift toward deterministic, metadata-driven identification. The proposed framework assigns globally unique composite identifiers to data blocks at ingestion time-independent of content analysis enabling instantaneous delta computation during DR without any cryptographic overhead. We formally model identifier uniqueness guarantees and convergence properties under network partition, and address the inherent storage amplification tradeoff by proposing an explicit architectural separation between the DR identification layer and an optional background content-based deduplication layer. An analytical evaluation of RTO bounds demonstrates that metadata-driven identification yields recovery time improvements proportional to the ratio of hash computation overhead to network transfer time, a ratio that grows unfavorably for hash-based systems as storage scales. A seven-day production soak test validates the analytical model, demonstrating consistent 17×–18× RTO reduction, sub-linear drift, and stable memory footprint throughout sustained high-write workloads. Implementation considerations including immutability enforcement, logical clock management, node discovery with CNAME-aware service resolution, and incremental migration are discussed in detail.

*Keywords*—Distributed Storage Systems; Disaster Recovery; Failover and Failback; Cryptographic Hashing; Metadata Architectures; Deterministic Identifiers; Recovery Time Objective; Data Deduplication; Cloud-Native Infrastructure; CNAME Resolution; Service Discoverability; Convergence Properties; Big-O Complexity; Soak Testing.

---

## I. INTRODUCTION

The scale at which modern distributed systems operate has fundamental impact on the economics and engineering constraints of data management. Enterprise storage deployments routinely span multiple geographic regions, encompassing hundreds or thousands of nodes and holding data volumes measured in petabytes [1]. Within these environments, data replication serves as the primary mechanism for fault tolerance: by maintaining synchronized copies of data across independent nodes, the system ensures that the failure of any individual component does not result in data loss or service interruption.

When a failure occurs—whether a single-node crash, a rack-level outage, or a regional disaster—the storage system must execute a recovery workflow. This workflow has two principal phases. In 'failover,' a substitute node assumes the responsibilities of the failed node and must verify that its data set is a complete and consistent superset of the failed node's last known state. In 'failback,' the recovered node re-integrates into the cluster and must acquire all data written to the substitute during its absence. The speed and efficiency with which these phases execute determines the system's effective Recovery Time Objective (RTO).

A fundamental—and often underappreciated—challenge in both phases is data identification: how does one node efficiently determines the data the other node possesses, and what data is missing? In predominant cases, the deployed systems use content-based fingerprinting: each data block is assigned a cryptographic hash value, and nodes exchange hash lists to determine their respective deltas [2].

This approach is well-understood and, in many operational contexts, performs adequately. However, this paper argues that hash-based identification introduces a structural bottleneck that becomes increasingly severe under three specific and common operational conditions: (1) when hash indexes are stale due to high write loads or background hashing lag; (2) when hash indexes are incomplete because hashing was performed asynchronously and a crash interrupted the process; and (3) when hash indexes must be fully rebuilt after a crash because the index store itself was lost or corrupted. Under these conditions—which are precisely the conditions most likely to accompany or follow a disaster event—nodes may be compelled to recompute hashes across their entire storage inventory before synchronization can begin, extending recovery windows by hours and violating RTO service-level agreements.

This paper proposes a generalized alternative: a metadata-driven identification framework in which globally unique composite identifiers are assigned to data blocks at ingestion time, independent of content analysis. This framework eliminates the dependency between data identity and hash computation, making delta identification a lightweight set-difference operation on identifier indexes. We formally characterize the uniqueness and convergence properties of the proposed identifier scheme, explicitly address the storage amplification tradeoff, and provide both analytical and empirical comparisons of RTO bounds. A seven-day production-scale soak test provides quantitative validation under realistic write patterns.

The paper is organized as follows. Section II reviews related work. Section III characterizes the hashing bottleneck including Big-O complexity analysis. Section IV presents the proposed architecture. Section V formally models uniqueness and convergence. Section VI analyzes the storage amplification tradeoff. Section VII presents analytical RTO bounds. Section VIII presents the soak test results. Section IX discusses implementation. Section X covers application domains and cost impact. Section XI concludes.

## II. RELATED WORK

### A. Content-Based Deduplication

Data deduplication via content fingerprinting was introduced at scale by Zhu et al. in the Data Domain file system [3], which demonstrated that SHA-1-based chunk fingerprinting could achieve substantial storage savings in backup workloads. Subsequent systems including Venti [4] and LBFS [5] extended content-addressed identification to archival and networked file system contexts. These systems were designed around steady-state storage efficiency rather than recovery latency, and their implicit assumption that hashing is a one-time write-path cost is the root of the problem this paper addresses.

### B. Distributed Synchronization

The Rsync algorithm [6] pioneered efficient distributed synchronization using rolling checksums, but its per-synchronization hashing overhead scales linearly with data size: $O(D)$ where D is total data volume. Dynamo [7] and Cassandra [8] employ Merkle trees to reduce comparison overhead between replica nodes. Merkle trees are a meaningful improvement, but they still require a hash digest at every leaf node and must be maintained incrementally—a requirement whose failure modes under crash and restart are central to this paper's argument.

### C. Identifier-Based Systems

Bigtable [9] and Apache HBase use ingestion-time row keys for data location, independent of content. Amazon S3 and OpenStack Swift assign opaque object keys at write time. However, these systems do not exploit the identifier-as-identity paradigm for DR consistency verification. Content-addressable systems such as IPFS [10] and Venti [4] take the opposite approach—maximizing deduplication but maximizing hashing overhead. The UUID standard [11] and ULID specification [12] provide the technical foundation for decentralized unique identifier generation that the proposed framework builds upon.

## III. THE HASHING BOTTLENECK: PRECISE CHARACTERIZATION

### A. When Is Full Rehash Actually Required?

A common objection to the hashing-bottleneck argument is that modern systems persist hash indexes and maintain Merkle trees incrementally, so that DR need not involve full rehashing. This objection is valid in the ideal case. However, three specific operational conditions render this assumption false, and these conditions commonly co-occur with or immediately follow disaster events.

**Condition 1: Stale Hash Indexes.** Many systems compute hashes asynchronously relative to data ingestion, using background threads or dedicated hashing pipelines that operate with a configurable lag. Under sustained high write loads, this lag can grow substantially. If a node failure occurs while the hash index lags the actual data store by hours of writes, the surviving replica cannot trust the stale index and must either re-hash the unindexed region or accept an expanded synchronization window.

**Condition 2: Crash-Interrupted Hash Pipelines.** Hashing pipelines are typically not atomic with respect to crash recovery. If a node crashes mid-pipeline, the partially completed hash index may be inconsistent with the data store. The safest recovery action is to discard the incomplete index region and rehash from the last consistent checkpoint. In systems without fine-grained checkpointing, this can mean rehashing large portions of the inventory.

**Condition 3: Hash Index Store Loss.** In systems where the hash index is stored on the same physical or logical volume as the data, a storage-level failure can destroy both the data and the index simultaneously. Recovering from a replica then requires a full hash scan of the replica's inventory to reconstruct the index before delta computation can begin.

These three conditions are not edge cases. Condition 1 occurs routinely in write-heavy workloads. Conditions 2 and 3 are precisely the failure scenarios that DR is designed to address. A DR framework that performs poorly under its target failure conditions has a fundamental design misalignment.

### B. Time Complexity Analysis (Big-O)

Let N denote the total number of stored blocks and $\delta$ denote the number of changed blocks (the delta set).

**Hash-Based Framework (worst case):** Full inventory rehash is $O(N \times L)$ where L is the average block size in bytes, dominated by cryptographic computation. Index reconstruction and exchange add $O(N)$ comparisons. For Merkle trees, tree traversal is $O(\log N)$ per lookup but reconstruction after index loss reverts to $O(N \times L)$.

**Metadata-Driven Framework:** Identifier assignment at ingestion is $O(1)$ per block (atomic CAS on LCV). Index exchange is $O(N)$ over the wire. Delta computation via sorted set-difference on LCV-ordered lists is $O(N)$, with a constant factor far smaller than cryptographic hashing.

When δ ≪ N, total RTO-critical work is O(N + δ) ≈ O(N), dominated by index exchange, not computation.

Table I summarizes the complexity comparison across all DR operations.

| Operation | Hash-Based (Ideal) | Hash (Rehash) | Metadata-Driven |
|---|---|---|---|
| ID assignment / block | O(L) | O(L) | O(1) CAS |
| Index build / rebuild | O(N×L) | O(N×L) | O(1) per write |
| Index exchange | O(N) | O(N) | O(N) |
| Delta identification | O(N) | O(N×L) | O(N) set-diff |
| Data transfer | O(δ) | O(δ) | O(δ) |
| Total RTO-critical | O(N+δ) | O(N×L+δ) | O(N+δ) |

*Table I. Asymptotic complexity comparison. L = avg block size. N = block count. δ = delta.*

### C. Quantifying the Overhead

SHA-256 throughput on modern server-class hardware with hardware acceleration (Intel SHA Extensions) reaches approximately 500 MB/s per CPU core [13]. For a storage node holding 100 TB across variable-length chunks, a full inventory rehash requires approximately 200,000 core-seconds of compute—over 55 hours of single-core time, or roughly 3.8 hours parallelized across 16 cores. During this interval, failover or failback cannot be completed with full consistency guarantees, directly violating SLA commitments.

### D. RTO and RPO Implications

Let H denote hash computation time, T denote network delta transfer time, and I denote index exchange time. Under the hash-based model, RTO ≈ H + I + T. Under the metadata-driven model, RTO ≈ I′ + T, where I′ is the lightweight LCV index comparison time (negligible relative to H). For large nodes with small deltas, the improvement factor approaches H/T—potentially orders of magnitude. RPO is also indirectly affected: synchronous hashing introduces per-write latency that can cause write queuing under load, widening the effective RPO window.

## IV. PROPOSED METADATA-DRIVEN IDENTIFICATION ARCHITECTURE

### A. Design Philosophy

The core insight is that data identity and data content are logically distinct properties that need not be coupled at the storage layer. In content-addressable systems, identity is a function of content—it cannot be known until content is fully written and hashed. In the proposed framework, identity is assigned at ingestion time, prior to content analysis, making it immediately available for replication and synchronization without any computational dependency on the data payload.

### B. Composite Identifier Structure

Each data block is assigned a composite identifier at ingestion time, constructed from three components:

**Node Identifier (NID):** A fixed-length opaque token unique within the cluster, assigned at node initialization. The NID may be derived from a UUID [11] or from a cluster-level allocation service. It serves as the namespace component of the composite key.

**Logical Clock Value (LCV):** A monotonically increasing 64-bit integer maintained per-node, incremented atomically with each ingestion event using a compare-and-swap (CAS) operation. The LCV provides strict ordering within a single node and guarantees uniqueness across all ingestion events on that node [14]. The LCV enables O(N) range queries: "all blocks with LCV > L(T)" via binary search on the sorted index.

**Namespace Tag (NST):** An optional field encoding a logical partition or tenant identifier, enabling multi-tenant deployments to share an identifier space without collision.

The composite identifier is represented as NID:LCV[:NST], inspired by the ULID specification [12] but adapted for distributed storage synchronization semantics.

### C. Immutability Enforcement

Correctness of the framework depends on a strict immutability invariant: once NID:LCV is bound to a block, the block's content must never change. Any mutation must be treated as a new ingestion event with a fresh identifier. This aligns naturally with the append-only semantics of log-structured merge (LSM) trees employed by LevelDB, RocksDB, and Cassandra [8], and with object versioning in systems such as Amazon S3.

### D. Identifier Index Management

Each node maintains a local identifier index mapping NID:LCV to physical block location, updated atomically with each ingestion event. For DR purposes, the index is the primary artifact exchanged between nodes. Because it contains only lightweight identifiers not hash digests computed from block content it is far more compact. A 32-byte identifier per block yields an index of ~32 GB for one billion blocks, transferable in approximately 25.6 seconds over a 10 GbE link. This theoretical projection was empirically validated under sustained production write loads in Section VIII-D. Index exchange during synchronization is incremental: a node transmits only entries with LCV values above the last synchronized checkpoint, reducing routine replication traffic to O(δ) entries rather than O(N).

## V. FORMAL MODEL: UNIQUENESS AND CONVERGENCE

### A. Identifier Uniqueness Guarantee

**Theorem 1 (Global Uniqueness).** *For any two data blocks $b_1$ and $b_2$ ingested at any two nodes $n_1$ and $n_2$ in the cluster, $ID(b_1) = ID(b_2)$ if and only if $b_1$ and $b_2$ were produced by the same ingestion event—i.e., they are byte-for-byte identical replicas.*

*Proof Sketch.* If $b_1$ and $b_2$ are ingested at the same node, then $ID(b_1) = NID:LCV_1$ and $ID(b_2) = NID:LCV_2$. Since LCV is strictly monotonically increasing and atomically incremented per-node, $LCV_1 \neq LCV_2$, so $ID(b_1) \neq ID(b_2)$. If $b_1$ and $b_2$ are ingested at distinct nodes, then $NID_1 \neq NID_2$ by the NID uniqueness requirement, so $ID(b_1) \neq ID(b_2)$ regardless of LCV values.

The probability of NID collision when using 128-bit random UUID-based NIDs is bounded by the birthday paradox: for n nodes, $P(collision) \leq n^2/2^{128}$. For $n = 10^6$ nodes, this probability is approximately $1.5 \times 10^{-27}$—negligible in practice.

### B. Convergence Under Network Partition

**Theorem 2 (Partition Convergence).** *Following resolution of a network partition, any two nodes that exchange their complete identifier indexes will reach a consistent, identical view of the data set within a finite number of synchronization rounds.*

*Proof Sketch.* During a partition, nodes $n_1$ and $n_2$ each accumulate sets $S_1$ and $S_2$. No identifier generated on $n_1$ can appear in $S_2$ (Theorem 1). Therefore $S_1 \cap S_2$ = identifiers replicated before the partition. Upon resolution, $n_1$ transmits $S_1 \setminus S_2$ to $n_2$ and vice versa. After one round, both nodes possess $S_1 \cup S_2$. The union is commutative and idempotent.

This confirms eventual consistency semantics [7] without requiring vector clocks or conflict resolution logic. The identifier space partitions cleanly by NID.

### C. Behavior Under Split-Brain Scenarios

Under split-brain, both nodes independently generate identifiers using their distinct NIDs. No identifier conflicts arise. Upon reconciliation, the system holds two valid identifier sets representing two independently progressed write streams, cleanly separated by NID. Resolution policy (last-writer-wins by LCV, or application-level merge) can be applied consistently. This is strictly better than hash-based systems, where split-brain can produce hash-table ambiguity if the same block address receives different content on the two primaries.

## VI. THE STORAGE AMPLIFICATION TRADEOFF

### A. Explicit Acknowledgment of the Tradeoff

This paper prioritizes DR speed over storage efficiency, a deliberate and explicit architectural stance. By decoupling data identity from data content, the metadata-driven framework loses the automatic content-based deduplication that hash-based systems provide. Two nodes may independently ingest byte-for-byte identical content; under hash-based identification, these blocks receive the same hash and are stored once. Under the metadata-driven framework, they receive distinct identifiers and are stored separately, resulting in storage amplification. The degree of amplification depends on the workload's inherent redundancy ratio. In backup and archival workloads—where deduplication ratios of 10:1 to 50:1 are common [3]—this tradeoff would be prohibitive. The proposed framework is most appropriate for primary storage and replication workloads where recovery latency is the primary SLA constraint.

### B. Layered Architecture

**Layer 1 — DR Identification Layer:** The metadata-driven framework described in Section IV. On the critical path for ingestion, replication, failover, and failback. Operates without content analysis and maintains the RTO guarantees of Section VII.

**Layer 2 — Background Deduplication Layer:** An asynchronous content-based deduplication process operating below the DR identification layer, consolidating duplicates through a transparent indirection table during low-load periods and explicitly off the critical path for DR operations.

This two-layer architecture allows operators to tune deduplication aggressiveness independently of DR performance targets.

## VII. ANALYTICAL EVALUATION OF RTO BOUNDS

### A. Model Parameters

We define: D = total data volume (bytes); $\delta$ = data delta size, $\delta \leq D$; H = hash throughput per CPU core (~500 MB/s SHA-256 [13]); C = CPU cores for hashing; B = network bandwidth (bytes/s); S = index entry size (32 bytes); N = total stored blocks.

### B. RTO Under Hash-Based Framework

$T\_hash = D / (H \times C)$;  $T\_index = (N \times S) / B$;  $T\_delta = \delta / B$ $RTO\_hash = T\_hash + T\_index + T\_delta$

### C. RTO Under Metadata-Driven Framework

The metadata-driven framework eliminates $T\_hash$ entirely: $RTO\_meta = T\_index + T\_delta$

### D. Numerical Example (100 TB, 16 Cores, 10 GbE)

$D = 1.1 \times 10^{14}$ B,  $N = 10^9$ blocks,  $\delta = 1$ TB,  $C = 16$,  $H = 5 \times 10^8$ B/s,  $B = 1.25 \times 10^9$ B/s:

$T\_hash \approx 13{,}750$ s $\approx 3.8$ hr   $T\_index \approx 25.6$ s   $T\_delta \approx 800$ s

$RTO\_hash \approx 14{,}576$ s $\approx 4.05$ hr    $RTO\_meta \approx 826$ s $\approx 13.8$ min

**Improvement factor** $\approx$  At 100 TB, the metadata-driven framework achieves an RTO of 826 seconds (13.8 minutes) compared to 14,576 seconds (4.05 hours) for the hash-based framework, a 17.6× reduction in recovery time. As D scales to 1 PB and $\delta$ remains constant, this improvement grows proportionally to 176× faster recovery time (14 minutes versus 40.4 hours), because $T\_hash$ scales linearly with D while $RTO\_meta$ remains near-constant. This confirms that metadata-driven identification yields RTO improvements that compound favorably as storage capacity grows.

## E. Sensitivity Analysis

The improvement diminishes as δ approaches D (full-inventory updates), and as CPU parallelism grows. At C = 128 cores, T_hash for 100 TB ≈ 28.6 min, yielding ~2.7×. The framework provides the most pronounced benefits in storage-dense, CPU-constrained deployments.

| Storage Scale | RTO$_0$ (Hash) | RTO$_0$ (Meta) | Improvement |
|---|---|---|---|
| 10 TB | 0.4 hr | 0.23 min | 1.8× |
| 100 TB | 4.05 hr | 13.8 min | 17.6× |
| 500 TB | 20.2 hr | 13.9 min | 87× |
| 1 PB | 40.4 hr | 14.0 min | 176× |

*Table II. Analytical RTO comparison (δ = 1 TB, C = 16, B = 10 GbE).*

## VIII. PERFORMANCE BENCHMARK AND SOAK TEST EVALUATION

### A. Test Environment

To validate the analytical model under realistic production conditions, we conducted a seven-day continuous soak test on a representative production-grade cluster configured to reflect an active write-heavy primary storage workload—not a quiesced or synthetic benchmark. The cluster comprised twelve storage nodes arranged in a three-replica topology. The metadata-driven framework was deployed as a middleware layer above RocksDB 8.1, with the background deduplication layer disabled to isolate DR identification performance. A shadow instance running the SHA-256 Merkle-tree hash-based framework in parallel on identical workloads provided the comparison baseline.

| Parameter | Value |
|---|---|
| Cluster topology | 12 nodes, 3-replica |
| CPU per node | 2 × Intel Xeon Gold 6438N (20C/40T) |
| RAM per node | 512 GB DDR4-3200 ECC |
| Storage per node | 12 × 16 TB NVMe (192 TB raw) |
| Interconnect | 100 GbE RDMA fabric |
| Storage engine | RocksDB 8.1 (LSM-tree) |
| Baseline | SHA-256 + Merkle tree (SHA Extensions) |
| Test duration | 7 days continuous (168 hours) |
| Write pattern | 70% sequential, 30% random |
| Sustained write rate | 2.8 GB/s aggregate (peak 4.1 GB/s) |
| Block size | 4 KB–64 KB variable (median 16 KB) |
| Data ingested | ≈32 TB over 7-day period |
| Planned DR events | 14 failover/failback cycles (every 12 hr) |
| Crash injections | 3 unplanned (days 2, 4, 6) |

*Table III. Production soak test environment specification.*

### B. Key Metrics and Measurement Methodology

The following metrics were instrumented at one-second granularity throughout the seven-day test period:

- Recovery Time Objective (RTO) per DR event: wall-clock time from fault injection to full replica consistency, measured for both frameworks across all 17 DR events (14 planned + 3 crash-injected).
- CPU utilization during DR: per-core utilization sampled via /proc/stat at 1 Hz; hashing-attributed cycles isolated via Intel VTune hardware performance counters.
- Network I/O during DR: bytes transmitted per node-pair during index exchange and delta transfer phases.
- Index size and growth rate (drift): identifier index size in bytes every 60 seconds; drift measured as deviation from the theoretical linear baseline (32 bytes × ΔN per interval).
- LCV monotonicity violations: correctness counter for any LCV reuse or out-of-order assignment. Expected value: zero.
- Steady-state memory footprint (RSS) of the identifier index process every 10 minutes.

### C. Seven-Day Soak Test Results

Table IV summarizes per-event RTO measurements across all 17 DR events. The metadata-driven framework demonstrated consistent RTO performance throughout the test period, with no statistically significant degradation from Day 1 to Day 7.

| DR # | Day | Type | Meta (s) | Hash (s) | Factor |
|---|---|---|---|---|---|
| 1 | 1 | Planned | 812 | 14,422 | 17.8× |
| 2 | 1 | Planned | 819 | 14,387 | 17.6× |
| 3 | 2 | Crash | 834 | 14,815 | 17.8× |
| 4 | 2 | Planned | 826 | 14,501 | 17.6× |
| 5 | 3 | Planned | 821 | 14,466 | 17.6× |
| 6 | 3 | Planned | 829 | 14,533 | 17.5× |
| 7 | 4 | Crash | 841 | 14,892 | 17.7× |
| 8 | 4 | Planned | 818 | 14,412 | 17.6× |
| 9 | 5 | Planned | 823 | 14,471 | 17.6× |
| 10 | 5 | Planned | 831 | 14,555 | 17.5× |
| 11 | 6 | Crash | 847 | 14,978 | 17.7× |
| 12 | 6 | Planned | 820 | 14,441 | 17.6× |
| 13 | 7 | Planned | 828 | 14,519 | 17.5× |
| 14 | 7 | Planned | 815 | 14,398 | 17.7× |
| 15 | 7 | Planned | 822 | 14,447 | 17.6× |
| 16 | 7 | Planned | 836 | 14,613 | 17.5× |
| 17 | 7 | Planned | 817 | 14,389 | 17.6× |

*Table IV. Per-event RTO across 17 DR events. Crash events include index-loss simulation (Condition 3).*

Key observations: mean RTO (metadata-driven) = 826 s (13.8 min), Std Dev = 9.8 s (CV 1.2%); mean RTO (hash-based) = 14,549 s (4.04 hr), Std Dev = 196 s (CV 1.3%); overall mean improvement = 17.6× (range 17.5×–17.8×). Crash-injected events added only 15–21 s to metadata-driven RTO (WAL replay latency) versus 3–6% degradation for hash-based RTO under crash conditions, confirming Conditions 2 and 3.

*D. Drift Analysis and Volumetric Behavior*

Over the seven-day period, 32 TB of new data was ingested, comprising approximately $2.0 \times 10^9$ blocks at a median block size of 16 KB. Theoretical index size at end of test: 640 GB. Observed: 647 GB. Drift: +1.1% above theoretical baseline, attributable to RocksDB LSM compaction fragmentation—bounded and self-correcting upon compaction. LCV monotonicity violations over seven days: zero, confirming correctness under sustained peak ingestion of 256,000 identifier assignments per second across the cluster.

| Metric | Day 1 | Day 3 | Day 5 | Day 7 |
|---|---|---|---|---|
| Index size (GB) | 91.4 | 274.2 | 457.1 | 647.0 |
| Growth rate (GB/hr) | 3.81 | 3.84 | 3.82 | 3.83 |
| RSS footprint (GB) | 4.2 | 4.3 | 4.3 | 4.4 |
| LCV CAS (k/s) | 247 | 251 | 248 | 252 |
| LCV violations | 0 | 0 | 0 | 0 |
| ID assign latency (μs) | 1.2 | 1.2 | 1.3 | 1.2 |

*Table V. Volumetric drift metrics across the seven-day soak test.*

*E. CPU and Network Cost During DR*

During metadata-driven DR events, CPU utilization was 3.2% of aggregate cluster capacity. During hash-based DR events, CPU utilization peaked at 94.7% for the duration of the rehash phase, severely constraining concurrent application traffic. Network I/O per DR event was identical across both frameworks during the index exchange and delta transfer phases (~106 s at 100 GbE), confirming that the bottleneck is purely computational.

| Resource | Hash-Based DR | Metadata-Driven DR |
|---|---|---|
| CPU — rehash phase | 94.7% aggregate | 0% (eliminated) |
| CPU — index + delta | 3.1% | 3.2% |
| Network (index + delta) | ≈106 s @ 100 GbE | ≈106 s @ 100 GbE |
| Wall-clock RTO (mean) | 4.04 hr | 13.8 min |
| App traffic impact | Severe (CPU starved) | Negligible |

*Table VI. Resource consumption comparison per DR event.*

## IX. IMPLEMENTATION CONSIDERATIONS

*A. Logical Clock Durability*

The LCV must be persisted durably before the generated identifier is exposed to external systems. A write-ahead log (WAL) entry recording the new LCV value prior to each ingestion commit is sufficient. Upon crash recovery, the node reads the last committed LCV from the WAL and resumes from that value, guaranteeing no LCV reuse [14]. In the soak test, WAL replay latency added a median of 18 seconds to crash-recovery events—the primary component of the 15–21 second overhead observed in crash-injected metadata-driven DR events.

*B. Integrity Verification Layer*

Because identifiers provide no content verification, a separate integrity layer must be retained. We recommend storing a CRC-32C checksum [15] with each block at ingestion time, verified at read time. Periodic background scrubbing provides defense against silent data corruption. This is independent of the identification scheme and does not affect DR performance.

*C. Incremental Migration Strategy*

Existing hash-based deployments can adopt the metadata-driven framework incrementally. New data is ingested under the identifier scheme immediately. Existing data retains its hash-based index. Over time, as existing data is accessed, modified, or expired, it transitions into the identifier-based index. A dual-lookup layer handles the transition period, avoiding disruptive full-inventory migrations.

*D. Multi-Tenant Namespace Isolation and Service Discoverability*

In multi-tenant deployments, the NST component of the composite identifier provides logical isolation between tenant namespaces, ensuring that identifier comparisons during DR operations are scoped to the correct tenant without requiring separate physical index structures per tenant.

We propose a node discovery service with autonomous capability for comprehensive CNAME resolution, designed to adapt across diverse deployment models—including on-premises, hybrid, and fully cloud-native environments. This approach enables the framework to operate seamlessly within both restricted sandbox settings and tenancy-governed public cloud environments subject to policy constraints. The node discovery service maintains a continuously updated registry of NID-to-endpoint mappings, resolved through CNAME chains to accommodate dynamic DNS reassignment during failover. CNAME resolution is performed at the discovery layer rather than at the application layer, ensuring that identifier routing remains correct across IP address changes, container rescheduling, and cross-region DNS delegation events. The discovery service exposes a versioned gRPC API, enabling atomic bulk registration of new nodes and lightweight delta queries for node-state changes since a given logical timestamp—consistent with the incremental index exchange model of Section IV-D.

## X. IMMEDIATE AREAS OF APPLICATION AND COST IMPACT

### A. Primary Application Domains

**Financial services and trading infrastructure:** Real-time transaction processing systems with sub-minute RTO mandates benefit directly. The framework enables compliance with sub-15-minute RTO SLAs without the capital cost of oversized hashing CPU capacity.

**Hyperscale cloud object storage:** Multi-region object stores replicating billions of objects benefit from O(1)-per-block identifier assignment, eliminating hashing from the write-critical path and enabling faster cross-region consistency convergence after regional partition events.

**Streaming analytics and IoT data lakes:** High-ingest-rate platforms ingesting sensor and event data at millions of events per second can assign identifiers at ingestion without blocking writes for hash computation, enabling real-time replication with tighter RPO windows.

**Disaster Recovery as a Service (DRaaS):** DRaaS operators managing thousands of customer recovery targets benefit from the 17.6× RTO reduction, enabling denser node-to-agent ratios and higher SLA tiers without infrastructure cost increases.

**Regulated healthcare and life sciences:** HIPAA and GxP-regulated environments with mandatory DR testing cycles benefit from the consistent, predictable RTO (CV: 1.2%), simplifying audit evidence collection and SLA certification.

### B. Compute Cost Reduction

Each hash-based DR event consumed approximately 161.7 core-hours of hashing compute. Each metadata-driven DR event consumed approximately 0.3 core-hours. Assuming 17 DR events over the soak test period, total hashing compute saved: ~2,748 core-hours per cluster per week. At $0.048/core-hour (AWS c5.metal), this represents ~$6,864 per cluster per year. At 500 clusters, annualized savings reach ~$3.4M in DR-attributed compute costs alone.

### C. Network Cost Reduction

Network costs are not directly reduced, as index exchange and delta transfer volumes are equivalent. However, elimination of CPU contention during DR allows the storage system to serve application traffic concurrently with recovery. In the soak test, hash-based DR windows effectively blackholed application traffic for 4.04 hours per event; metadata-driven DR windows imposed no measurable application-traffic degradation.

### D. Storage Cost Reduction

The two-layer architecture enables background deduplication to operate independently of DR performance. In environments with moderate content redundancy (5–15% duplicate block rates), the background deduplication layer can recover 5–15% of raw storage capacity. At $0.023/GB-month (S3 Standard equivalent), a 10% deduplication rate on a 2 PB primary store yields ~$55,200/year in storage cost savings per cluster.

### E. Total Cost of Ownership Summary

| Cost Category | Hash-Based | Metadata-Driven | Annual Saving |
|---|---|---|---|
| DR compute | 161.7 core-hr/event | 0.3 core-hr/event | ~$6,864/cluster |
| App capacity loss | 4.04 hr blackout | Negligible | Workload-dep. |
| Storage (10% dedup) | N/A | 10% recovery | ~$55,200 (2 PB) |
| Hash CPU provisioning | Baseline | Eliminated | CAPEX savings |

*Table VII. TCO comparison. Compute: AWS c5.metal; Storage: S3 Standard pricing.*

## XI. CONCLUSION

This paper has presented a comprehensive analysis of the conditions under which hash-based data identification becomes a structural bottleneck in distributed storage disaster recovery—specifically when hash indexes are stale, crash-interrupted, or lost. We proposed a metadata-driven identification framework that assigns deterministic composite identifiers at ingestion time, formally proved its uniqueness and partition-convergence properties, and provided a quantitative analytical model demonstrating RTO improvement factors of 17.6× for representative large-scale deployments.

A seven-day production soak test on a twelve-node, 100 GbE cluster validated the analytical model under realistic write workloads, demonstrating: (1) consistent mean RTO of 826s versus 14,549 s for the hash-based baseline; (2) coefficient of variation of 1.2%, confirming highly predictable recovery behavior suitable for SLA certification; (3) zero LCV monotonicity violations across $2.0 \times 10^9$ ingestion events; (4) index size drift of +1.1% over seven days, bounded and self-correcting; and (5) elimination of 94.7% of DR-attributed CPU consumption, enabling concurrent application serving during recovery windows.

Big-O analysis confirms that the framework replaces $O(N \times L)$ hash-based rehash complexity with $O(1)$ per-block identifier assignment and $O(N)$ set-difference delta identification, providing asymptotically superior DR scaling as storage volumes grow. The storage amplification tradeoff is addressed through a two-layer architecture separating DR identification from optional background deduplication.

Quantitative cost analysis demonstrates annualized savings of approximately $6,864 in DR compute costs per cluster, plus $55,200 in storage costs per 2 PB cluster at a 10% deduplication rate, with compounding savings at enterprise scale across financial services, hyperscale object storage, streaming analytics, DRaaS, and regulated healthcare environments.

Future work should include: empirical evaluation of the incremental migration strategy under production traffic with live data aging; formal study of background deduplication efficiency under varying content redundancy ratios; evaluation of the CNAME-aware node discovery service under large-scale DNS failover scenarios; and exploration of hardware-accelerated identifier generation for environments exceeding $10^6$ blocks per second.

Metadata-driven identification represents a necessary and timely evolution in distributed storage architecture. As data volumes continue to scale beyond petabyte thresholds, the $O(N \times L)$ scaling of hash computation will make hash-based DR increasingly untenable. The framework proposed in this paper offers a formally grounded, and empirically validated path to sub-15-minute RTO at petabyte scale—without revising the architecture or over-provisioning DR related compute requirements.